\documentclass[prx,twocolumn,aps,longbibliography]{revtex4-1}
\usepackage{amsmath,amsfonts,graphicx,enumerate,color}
\usepackage{bm}
\usepackage{tabularx}
\usepackage{amssymb}
\usepackage{amsthm}
\usepackage{mathrsfs}
\newcommand{\pa}{\partial}

\newcommand{\B}[1]{{\bm{#1}}}
\newcommand{\C}[1]{{\mathcal{#1}}}

\begin{document}

\title{Failing softly: A fracture theory of highly-deformable materials}

\author{Tamar Goldman Bou\'e$^1$, Roi Harpaz$^2$, Jay Fineberg$^1$ and Eran Bouchbinder$^2$}
\address{$^1$Racah Institute of Physics, Hebrew University of Jerusalem, Jerusalem 91904, Israel\\
$^2$Chemical Physics Department, Weizmann Institute of Science, Rehovot 7610001, Israel}

\date{\today}
\begin{abstract}
Highly-deformable materials, from synthetic hydrogels to biological tissues, are becoming increasingly important from both fundamental and practical perspectives. Their mechanical behaviors, in particular the dynamics of crack propagation during failure, are not yet fully understood. Here we propose a theoretical framework for the dynamic fracture of highly-deformable materials, in which the effects of a dynamic crack are treated with respect to the nonlinearly deformed (pre-stressed/strained), non-cracked, state of the material. Within this framework, we derive analytic and semi-analytic solutions for the near-tip deformation fields and energy release rates of dynamic cracks propagating in incompressible neo-Hookean solids under biaxial and uniaxial loading. We show that moderately large pre-stressing has a marked effect on the stress fields surrounding a crack's tip. We verify these predictions by performing extensive experiments on the fracture of soft brittle elastomers over a range of loading levels and propagation velocities, showing that the newly developed framework offers significantly better approximations to the measurements than standard approaches at moderately large levels of external loadings and high propagation velocities. This framework should be relevant to the failure analysis of soft and tough, yet brittle, materials.
\end{abstract}

\maketitle

\section{Background and motivation}
\label{intro}

Material failure is mediated by the propagation of cracks, which are spatially-extended dissipative defects that concentrate large deformation and stresses near their tips. The latter is an essential physical property of cracks that highlights their basic role in material failure: cracks strongly amplify remotely applied stresses near their tips, precipitating fracture as they propagate.

The classic approach to the fracture of brittle materials, Linear Elastic Fracture Mechanics (LEFM) \cite{Lawn.93,Freund.90,99Bro}, is a perturbative approach that treats all relevant physical quantities to linear order in the elastic deformation with respect to the undeformed state of the material. Here, all nonlinearities and dissipation are assumed to be confined to a microscopically small region near the crack tip, and are neglected.

LEFM does not treat separately the effects of the externally applied loading (pre-stress/strain) and the presence of the crack itself, due to the linearity assumption. This physical picture makes sense as long the materials of interest fail when the applied forces are small and the deformation is indeed linear elastic (i.e. infinitesimal) everywhere except for the immediate vicinity of the crack tip. This has always been the case in ``traditional'' brittle materials such as glasses, ceramics and brittle polymers, where failure occurs at applied strains of the order of $1\%$.

Recently, an extended theory of dynamic fracture -- the ``Weakly Nonlinear Theory of Dynamic Fracture'' -- was developed \cite{Bouchbinder.08a,Livne.08,Bouchbinder.09,Autonomy,Bouchbinder.2014}. This theory still treats the large scales of a fracture problem as linear elastic, but explicitly takes into account the leading order nonlinear elastic corrections near the tip of a crack. In the latter region, strains are necessarily large and nonlinearities of the elastic fields are important for understanding the physical state that actually drives material failure.

The weakly nonlinear theory introduced a new intrinsic length scale (i.e. independent of the external geometry and crack's dimensions) that emerges from the competition between linear and weakly nonlinear deformation. The new length scale, which is missing in LEFM and characterizes the scale near the crack tip where LEFM breaks down, has been argued to play an important role in determining the crack's path \cite{Bouchbinder.2009a} and has been shown to play a decisive role in crack instabilities \cite{Baumberger2008, Baumberger2010, Ronsin2014,Goldman.12}. This theory has recently received significant experimental support \cite{Goldman.12,Goldman.15}.

Of late, classes of materials in which large deformations precede failure have become increasingly important and have been extensively studied in various contexts. These highly-deformable materials include synthetic elastomers, biopolymers, gels and a broad range of soft biological tissues (e.g. arterial walls, veins, skin, tendons etc.). Reviewing  even a small portion of the work done on the fracture of soft, highly-deformable, materials --- which encompass various scientific disciplines and communities --- goes well beyond the scope of this paper. We can only mention here a few examples.

The fracture resistance of some synthetic materials, such as double-network hydrogels made of ionically and covalently crosslinked networks \cite{Gong.2003, Gong.2010,Sun.2012, Tetsuharu.2013, Vlassak.2014, Leibler.2014}, can be made to be enormous and new applications are expected to abound. Basic phenomena such as delayed fracture of soft solids \cite{Bonn1998}, and surface tension-induced and capillary fracture of gels \cite{Daniels2007, Spandagos2012a, Spandagos2012b, Bostwick2013} have been studied. Finally, the work of Baumberger and coworkers on the quasi-static fracture of soft materials has both elucidated the fracture mechanisms of biopolymer gels \cite{Baumberger2006a, Baumberger2006b, Baumberger2009} and established the existence of an intrinsic length scale associated with elastic nonlinearities \cite{Baumberger2008, Baumberger2010, Ronsin2014}.

It is important to note that many of the materials of interest here, although highly-deformable and tough, are still brittle. The materials that we are considering here are materials where high-deformability is dominated by nonlinear elastic behavior, with little ductility. This type of behavior is typical in many of the new tough elastomers that are being developed. Since the bulk dissipation can be ignored, these materials are indeed brittle -- as the only dissipation takes place in the near vicinity of the crack tip.

When failure occurs under the application of large background deformation (pre-stress/strain), the deformation is nonlinear everywhere in the material, invalidating the assumption of remote linear elastic deformation. To address this problem, some works considered the fully nonlinear field equations for certain classes of highly-deformable materials and non-perturbatively derived the leading asymptotic fields in the inner most crack tip region. The vast majority of these works focussed on static cracks \cite{Knowles.73, 82Steph, LeStumpf.93, 95Geubelle, Long.2011}, though not all \cite{Tarantino.99, 05Tarantino, Livne.2010}. Such an asymptotic approach would be useful if the asymptotic fields are generic and universally linked to the remote loading, something which has not yet been established. Marder, following a different approach combining numerical and analytic techniques, developed a rather comprehensive dynamic fracture theory of rubber \cite{Marder.JMPS.2006}. Despite these important efforts, we are still far from having a well-established general theory of the dynamic fracture of highly-deformable, strongly nonlinear, materials.

In this paper, we develop a theoretical framework in which the background, possibly finite (nonlinear), deformation induced by the external loading in the absence of a crack (pre-stress/strain) is treated non-perturbatively. Then, the effects of the crack on its near-tip region are treated perturbatively to {\em second order} with respect to the background deformation. This theory will be shown to offer quantitatively good approximations to experimental data at moderately large background deformation. Consequently, we occasionally refer to it as the moderately large deformation theory.

The development of a perturbative fracture theory in the presence of non-infinitesimal background deformation is conceptually non-trivial. It raises the following question: if the background deformation is finite (nonlinear) and a crack significantly amplifies the background (remotely applied) deformation near its tip, is it justified to treat the latter as a relatively small correction to the former?

To quantitatively address this issue, we first develop the moderately large deformation theory and derive its near crack tip solutions (both analytically and semi-analytically). We then systematically and quantitatively compare it to the weakly nonlinear theory. Both theories are also compared to extensive direct measurements of the near-tip deformation fields of dynamic cracks propagating in a brittle elastomer gel, where the background deformation is increased in a controlled manner.

Our results show that while the two theories agree with each other at relatively small background deformation, the moderately large deformation theory offers significantly better approximations to the experimental data at moderately large levels of external loadings (pre-stress/strain) and high propagation velocities.

Theoretical frameworks that invoke perturbations of a pre-stressed state -- which are sometimes referred to as ``Mechanics of incremental deformation'' -- are not new in themselves \cite{Biot.1965,Guz.1999}. A classical example is the theory of small amplitude waves in nonlinearly pre-stressed materials, which has recently attracted renewed attention \cite{Destrade.2007}. In this case, a perturbative approach relative to the pre-stressed/strained state of the material is fully justified as the wave amplitude can remain small relative to the large background deformation. As explained above, this is not automatically the case in fracture dynamics where the background deformation is significantly amplified near the crack tip and hence in principle may not be treated as a small perturbation.

A number of authors have previously discussed such approaches to fracture~\cite{Guz.1999, Brock_Hanson}, where the effect of the crack was treated perturbatively to {\em linear order} with respect to the background deformation. As far as we can tell, however, these authors did not address at all the range of validity of the approach. In particular, their perturbative approach was confined to linear order, which -- as we show below -- is insufficient since higher order effects (in particular, weakly nonlinear effects with respect to the background deformation) play an important role.

Furthermore, to the best of our knowledge, quantitative comparisons of theoretical predictions to detailed experimental data, as we do here, have not previously been performed. We believe that the combined theoretical-experimental results presented in this paper offer a useful framework to quantitatively address the fracture properties and dynamics of highly-deformable materials.

\section{Theoretical framework}
\label{theory}

To lay down the theoretical grounds for the approach we propose, consider a dynamic crack propagating in a 2D nonlinear elastic solid described by an energy functional $U(\B F)$. The deformation gradient tensor $\B F$ is defined as $\B F(\B x,t)\!=\!\nabla\!_{\B x} \B \varphi(\B x,t)$, where the motion $\B \varphi(\B x,t)$ is a continuous, differentiable and invertible mapping between a reference (undeformed) configuration described by $\B x$ and a deformed configuration described by $\B x'$, such that $\B x'\!=\!\B \varphi(\B x,t) \!=\! \B x + {\B u}(\B x, t)$. ${\B u}(\B x, t)$ is the displacement vector field. Linear momentum balance can be expressed in the reference configuration as \cite{Holzapfel}
\begin{equation}
\nabla\!_{\B x}\!\cdot\!\B s = \rho\,\ddot{\B \varphi} \ ,
\label{momentum}
\end{equation}
where $\B s$ is the first Piola-Kirchhoff stress tensor, $\B s\!=\!\pa_{\B F}U(\B F)$, and $\rho$ is the time-independent reference mass density.

The crack is assumed to follow a straight trajectory and to propagate steadily at a velocity $v$ along the positive $x$-axis. Under symmetric tensile loading along the $y$-axis, that is under mode-I fracture conditions, the crack faces are being separated and hence are traction-free. These traction-free boundary conditions on the crack faces can be expressed in the undeformed configuration as \cite{Holzapfel}
\begin{equation}
s_{xy}(r,\theta=\pm\pi)=s_{yy}(r,\theta=\pm\pi)=0 \ ,
\label{generalBC}
\end{equation}
where $(r,\theta)$ is a polar coordinates system co-moving with the crack tip ($r\!=\!0$ is the tip location and $\theta\!=\!0$ is the propagation direction).

The solution of Eq. \eqref{momentum}, with the boundary conditions of Eq. \eqref{generalBC} and a constitutive relation $\B s\!=\!\pa_{\B F}U(\B F)$ for a general nonlinear energy functional $U(\B F)$, is analytically intractable. To make progress, some approximations are invoked, most notably in situations in which fracture occurs under small background deformation. In such cases, the displacement gradient tensor $\B H\!\equiv\!\nabla\!_{\B x} \B u$ is treated as small everywhere except for a small zone around the crack tip, and a perturbative approach is developed. LEFM \cite{Lawn.93,Freund.90,99Bro} and the weakly nonlinear theory of fracture \cite{Bouchbinder.08a,Livne.08,Bouchbinder.09,Autonomy,Bouchbinder.2014} fall under this category. Highly-deformable materials, on the other hand, fail at large, nonlinear, background deformation and in principle the status of a perturbative approach is not clear.

\subsection{General formulation}

\begin{figure}[here]
\centering\includegraphics[width=0.39\textwidth]{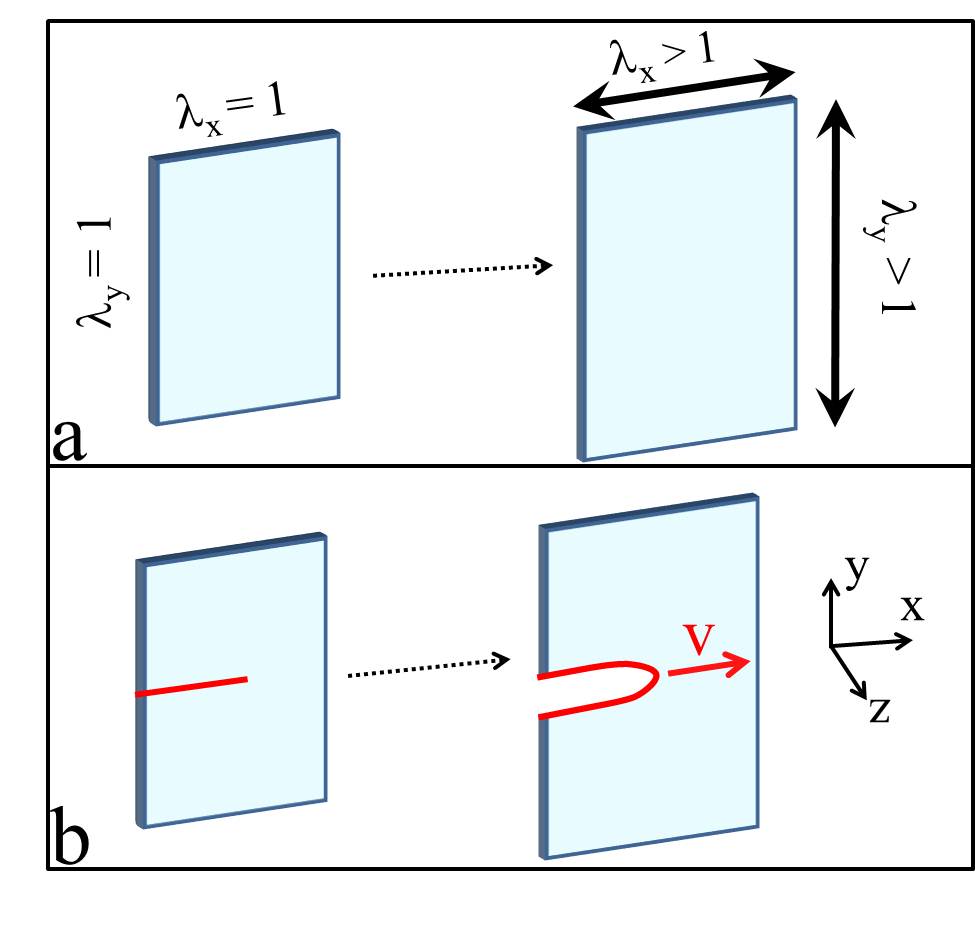}
\caption{A sketch that illustrates the decomposition in Eq.~\eqref{phi_decom}. (a) An undeformed configuration is uniformly stretched in two directions, described by two principal stretches $\lambda_{x,y}\!>\!1$, corresponding to the first terms on the right-hand-side of Eq.~\eqref{phi_decom}. The effect of the crack relative to this pre-stretched state is described by the field ${\B {\C U}}(x,y,t)$. (b) The actual motion is described by ${\B \varphi}(x,y,t)$ which takes into account both the presence of the crack and the external stretch. On the left, the undeformed configuration with an unopened crack is shown. On the right, the stretched configuration with an opened crack propagating at a velocity $v$ is shown. The coordinate system showing the propagation direction ($x$), the tensile loading direction ($y$) and the thickness direction ($z$) is added for clarity.}\label{fig1}
\end{figure}

To develop our approach, we write the motion $\B \varphi(x,y,t)$ as
\begin{eqnarray}
\varphi_x(x,y,t) &=& \lambda_x\,x + {\C U}_x(x,y,t) \ ,\nonumber\\
\varphi_y(x,y,t) &=& \lambda_y\,y + {\C U}_y(x,y,t) \ .
\label{phi_decom}
\end{eqnarray}
Here we decompose the total motion into a contribution emerging from the external loading $\lambda_{x,y}$ in the absence of a crack (the pre-stress/strain), which are the stretches in the $x$ and $y$ directions respectively, and into the effect of the crack quantified by $\B {\C U}(x,y,t)$. For simplicity, and to later allow direct comparison with experiments, we assume hereafter that $\lambda_{x,y}$ are constants. When $\lambda_{x}\!=\!\lambda_y\!=\!1$, $\B {\C U}(\B x, t)$ becomes the ordinary displacement field $\B u(\B x, t)$. In this case, LEFM corresponds to the linear approximation in the displacement gradient $\B H$ and the weakly nonlinear theory to the second order approximation in $\B H$ (leading order nonlinearity) \cite{Bouchbinder.08a,Livne.08,Bouchbinder.09,Autonomy,Bouchbinder.2014}. The decomposition in Eq.~\eqref{phi_decom} is sketched in Fig.~\ref{fig1}.

We now consider situations where $\lambda_{x,y}$ deviate to a finite degree from unity. When the motion $\B \varphi$ of Eq. \eqref{phi_decom} is substituted in $\B s\!=\!\pa_{\B F}U(\B F)$, which is then substituted in Eq. \eqref{momentum}, a nonlinear set of equations in $\B {\C U}$ is obtained. In this paper we solve this problem perturbatively to second order in $\nabla\!_{\B x}{\B {\C U}}$, i.e. employ the expansion
\begin{equation}
\B {\C U} \simeq \B {\C U}^{(1)} + \B {\C U}^{(2)}
\label{SOLexpansion}
\end{equation}
in the near tip vicinity of a dynamic crack, which is understood as a displacement gradients expansion.

The expansion in Eq. \eqref{SOLexpansion} corresponds to an expansion of the stress of the form $\B s \simeq {\B s}^{(0)}+{\B s}^{(1)}+{\B s}^{(2)}$, where ${\B s}^{(0)}$ simply corresponds to the pre-stress. Using the latter and transforming into a frame of reference moving with the crack tip, momentum balance in Eq. \eqref{momentum} translates into two linear problems which take the form
\begin{eqnarray}
\label{1st_2nd_a}
&&\nabla\!_{\B x}\!\cdot\!\B s^{(1)} = {\B {\C L}}[\B {\C U}^{(1)}] = \rho v^2 \pa_{xx} \B {\C U}^{(1)} \ ,\\
&&\nabla\!_{\B x}\!\cdot\!\B s^{(2)} = {\B {\C L}}[\B {\C U}^{(2)}]+{\B {\C F}}[\B {\C U}^{(1)}] = \rho v^2 \pa_{xx} \B {\C U}^{(2)} \ ,
\label{1st_2nd_b}
\end{eqnarray}
and Eq. \eqref{generalBC} gives rise to the corresponding boundary conditions on $\B {\C U}^{(1)}$ and $\B {\C U}^{(2)}$. ${\B {\C L}}[\,\cdot\,]$ is a vectorial linear differential operator, which depends on the nonlinear energy functional $U(\B F)$ and $\lambda_{x,y}$, and has the form of an effective anisotropic linear elastic operator. Anisotropy here emerges due to nonlinearity in $U(\B F)$ and the possibly symmetry-breaking external stretches $\lambda_{x,y}$. Since ${\B {\C L}}[\,\cdot\,]$ is a linear elastic operator, we expect the asymptotic near tip solution of Eq.~\eqref{1st_2nd_a} to give rise to the standard singularity $\nabla\!_{\B x}{\B {\C U}}^{(1)}\!\sim\!1/\sqrt{r}$ at small $r$. This square root singularity implies a parabolic crack tip opening profile \cite{Freund.90}.

What do we expect to occur to second order? We are guided by the intuition gained by solving the weakly nonlinear problem for infinitesimal deformations \cite{Bouchbinder.08a,Livne.08,Bouchbinder.09,Autonomy,Bouchbinder.2014}. Hence, the second order problem, i.e. Eq.~\eqref{1st_2nd_b}, features the same linear operator ${\B {\C L}}[\,\cdot\,]$ as in the first order problem, but also an effective body force ${\B {\C F}}[\B {\C U}^{(1)}]$ corresponding to quadratic contributions emerging from the first order solution. In particular, we have ${\B {\C F}}[\B {\C U}^{(1)}]\!\sim\!\nabla\!_{\B x}[\nabla\!_{\B x}{\B {\C U}}^{(1)}]^2\!\sim\!1/r^2$. The boundary condition which ${\B {\C U}}^{(2)}$ satisfies, emerging from a consistent expansion of Eq. \eqref{generalBC}, features an effective surface force proportional to $1/r$. As in the weakly nonlinear theory derived for small background strains, this structure is expected to lead to $\nabla\!_{\B x}{\B {\C U}}^{(2)}\!\sim\!1/r$, i.e. to a stronger singularity than the linear problem. ${\B {\C U}}^{(2)}$ is expected to modify the crack tip shape and to introduce a new lengthscale into the problem~\cite{Bouchbinder.08a,Livne.08,Bouchbinder.09,Autonomy,Bouchbinder.2014}, the length at which $|\nabla\!_{\B x}{\B {\C U}}^{(1)}|\!\simeq\!|\nabla\!_{\B x}{\B {\C U}}^{(2)}|$.

The elastic fields transport a finite amount of energy into the tip region, as quantified by the J-integral \cite{Freund.90}
\begin{equation}
J\!=\!\int_{\C C}\!\left[\left(U(\B F)\!+\!\tfrac{1}{2}\rho\,\left[\pa_t \varphi_i\right]^2\right)v\,n_x + s_{ij}\,n_j\,\pa_t \varphi_i\right]\!d{\C C}\ ,
\label{Gint}
\end{equation}
where ${\C C}$ is a contour encircling the tip and $\B n$ is an outward unit vector on ${\C C}$. This integral is path-independent for steady-state crack propagation and for any contour ${\C C}$ within a non-dissipative region described constitutively by the elastic energy functional $U(\B F)$. $G(v)\!=\!J/v$, the energy release rate whose dimensions are energy per unit crack area, is dissipated near the tip. This dissipation is quantified by the fracture energy $\Gamma(v)$ -- a measure of the material's resistance to crack propagation -- which is a fundamental material function assumed to depend only on the crack propagation velocity. Energy balance implies that $G(v)\!=\!\Gamma(v)$, which enables us to use $G(v)$ to calculate $\Gamma(v)$ (see below).

The approximate solution for the motion $\B \varphi(x,y)$ depends in a nontrivial way on the nonlinear energy functional $U(\B F)$ and on the background stretches $\lambda_{x,y}$. In the next subsections we will demonstrate how to actually derive the solution and explore some of its physical properties.

\subsection{Analytic example of first order asymptotic fields}
\label{subsec:analytic}

To see how all of this works, we discuss an explicit example that can be worked out analytically in a rather straightforward manner. We consider an incompressible neo-Hookean material under plane-stress conditions, whose nonlinear energy functional takes the form \cite{Knowles.83}
\begin{equation}
U(\B F) = \frac{\mu}{2}\left(F_{ij}F_{ij}+\Delta_z^2-3\right) \ .
\label{NH}
\end{equation}
Here $\mu$ is the shear modulus, $\B F$ is the 2D deformation gradient and $\Delta_z(x,y)\!=\![\hbox{det}(\B F)]^{-1}\!=\![\pa_x \varphi_x\,\pa_y \varphi_y-\pa_x \varphi_y \,\pa_y \varphi_x]^{-1}$ is the out-of-plane stretch. We choose this energy functional because it is relevant for many highly-deformable materials and it will allow us later to compare our predictions to direct experimental measurements.

The stress tensor $\B s$ corresponding to $U(\B F)$ in Eq. \eqref{NH} reads $s_{ij}\!=\!\mu\left(\pa_j\varphi_i\!-\!\Delta_z^3\,\epsilon_{ik}\,\epsilon_{jl}\,\pa_l\varphi_k \right)$,
where $\epsilon_{ij}$ is the 2D alternator (i.e. $\epsilon_{xx}\!=\!\epsilon_{yy}\!=\!0$, $\epsilon_{xy}\!=\!-\epsilon_{yx}\!=\!1$).
The momentum balance of Eq. \eqref{momentum} takes the form
\begin{align}
&\mu\nabla^2\!\varphi_x+\mu\left[\pa_y\Delta_z^3\,\pa_x \varphi_y-\pa_x\Delta_z^3\,\pa_y \varphi_y \right]=\rho\,\ddot{\varphi}_x\ ,\nonumber\\
&\mu\nabla^2\!\varphi_y+\mu\left[\pa_x\Delta_z^3\,\pa_y \varphi_x-\pa_y\Delta_z^3\,\pa_x \varphi_x \right] = \rho\,\ddot{\varphi}_y \ ,
\label{NH_EOM}
\end{align}
while the traction-free boundary conditions of Eq. \eqref{generalBC} read
\begin{align}
&s_{xy}(r,\theta\!=\!\pm\pi)=\mu(\pa_y\varphi_x+\Delta_z^3\,\pa_x\varphi_y)|_{\theta=\pm\pi}=0 \ ,\nonumber\\ &s_{yy}(r,\theta\!=\!\pm\pi)=\mu(\pa_y\varphi_y-\Delta_z^3\,\pa_x\varphi_x)|_{\theta=\pm\pi}=0 \ .
\label{NH_BC}
\end{align}

We focus now on pre-stressing/straining of the form $\lambda_x\!=\!\lambda_y\!=\!\lambda$, i.e. on a biaxial stretch, which is simpler than uniaxial stretching due to its symmetry. This loading corresponds to a pre-stress of the form $s^{(0)}_{ij}\!=\!(\lambda-\lambda^{-5})\mu\,\delta_{ij}$. Substituting Eqs. \eqref{phi_decom}-\eqref{SOLexpansion} in Eqs. \eqref{NH_EOM}, we obtain to linear order (again in the co-moving frame of reference)
\begin{equation}
\mu\nabla^2{\B {\C U}^{(1)}}+ 3\lambda^{-6}\mu \nabla(\nabla\cdot{\B {\C U}^{(1)}})=\rho\,v^2\pa_{xx}{\B {\C U}}^{(1)} \ ,
\label{1st}
\end{equation}
which has the structure Eq.~\eqref{1st_2nd_a}. Using Eqs. \eqref{NH_BC} we obtain the following boundary conditions at $\theta\!=\!\pm\pi$
\begin{eqnarray}
&\mu\,\pa_y {\C U}^{(1)}_x + \lambda^{-6}\mu \,\pa_x {\C U}^{(1)}_y = 0\ ,\nonumber\\
&(1+ 3\lambda^{-6})\mu \,\pa_{y} {\C U}^{(1)}_y  + 2\lambda^{-6}\mu\, \pa_x {\C U}^{(1)}_x = 0 \ .
\label{1st_BC}
\end{eqnarray}
Obviously, LEFM is recovered in the limit $\lambda\!\to\!1$ (as the material is incompressible, the resulting LEFM plane-stress problem involves a Poisson's ratio of $\tfrac{1}{2}$).
Equation \eqref{1st} has the structure of an isotropic linear elastic problem (Lam\'e equation) with a $\lambda$-independent shear modulus and a first Lam\'e coefficient of the form
$(3\lambda^{-6}\!-\!1)\mu$. Consequently, we can employ rather standard complex functions techniques \cite{Freund.90,99Bro} to obtain the following asymptotic (near tip) analytic solution
\newpage
\begin{widetext}
\begin{eqnarray}
{\C U}^{(1)}_x(r,\theta; v, \lambda) &=& \frac{2K_I \sqrt{r}}{\mu \sqrt{2\pi}  D(v,\lambda)}\left[(\lambda^{-6}+\alpha_s^2) \sqrt{\gamma_d}\cos{\left(\tfrac{\theta_d}{2}\right)}-
(1+\lambda^{-6})\alpha_d \alpha_s \sqrt{\gamma_s}\cos{\left(\tfrac{\theta_s}{2}\right)}\right]\ ,\nonumber\\
{\C U}^{(1)}_y(r,\theta; v, \lambda) &=& -\frac{2K_I \sqrt{r}\,\alpha_d}{\mu \sqrt{2\pi}  D(v,\lambda)}\left[(\lambda^{-6}+\alpha_s^2 )\sqrt{\gamma_d}\sin{\left(\tfrac{\theta_d}{2}\right)}-
(1+\lambda^{-6})\sqrt{\gamma_s}\sin{\left(\tfrac{\theta_s}{2}\right)} \right] \ .
\label{biaxial}
\end{eqnarray}
\end{widetext}

The quantities $\alpha_{s,d}$, $\gamma_{s,d}$ and $\theta_{s,d}$ are analogous to their standard LEFM counterparts, rather with a pre-stretch dependent dilatational wave-speed $c_d(\lambda)\!=\!\sqrt{1+3\lambda^{-6}}c_s$, where the shear wave-speed $c_s\!=\!\sqrt{\mu/\rho}$ is unaffected by $\lambda$. In particular, $\alpha_{s,d}\!=\!\sqrt{1\!-\!(v/c_{s,d})^2}$, $\gamma_{s,d}\!=\! \sqrt{1\!-\!(v\sin{\theta}/c_{s,d})^2}$ and $\tan({\theta_{s,d}})\!=\!\alpha_{s,d}\tan({\theta})$. The pre-stretch dependent analog of the Rayleigh function takes the form $D(v,\lambda)\!=\! 2(1+\lambda^{-6})\alpha_d\alpha_s-\left(1+\alpha^2_s \right)\left(\lambda^{-6}+\alpha^2_s\right)$ and $K_I$ is the mode-I stress-intensity-factor \cite{Freund.90,99Bro}. The solution in Eq. \eqref{biaxial} features the expected singularity, $\nabla\!_{\B x}{\B {\C U}}^{(1)}\!\sim\!1/\sqrt{r}$, where the standard LEFM solution~\cite{Freund.90} is recovered in the limit $\lambda\!\to\!1$.

The analytic solution in Eq. \eqref{biaxial} has several physical implications. First, it can be used to calculate the J-integral of Eq. \eqref{Gint}, employing the linear elastic approximation of $U(\B F)$, yielding
\begin{equation}
G(v)\!=\!\frac{v^2 \alpha_d K_I^2(v,\lambda)}{2\,c_s^2 D^2(v,\lambda) \mu}\left[(1+\lambda^{-6})^2\alpha_d\alpha_s - (\lambda^{-6}+\alpha_s^2)^2 \right]\ .
\label{G}
\end{equation}
The stress-intensity-factor $K_I$ cannot be obtained from the asymptotic solution, rather from the global boundary value problem. It can be calculated analytically only in relatively simple cases and in general it is obtained numerically or measured experimentally. Once it is available, Eq. \eqref{G} allows one to calculate the fracture energy $\Gamma(v)$, a basic material property, through the relation $\Gamma(v)\!=\!G(v)$.

Alternatively, if $\Gamma(v)$ is known (either from a proper dissipation theory, which is very rare, or through independent measurements) one can calculate $K_I(v,\lambda)$ using energy balance and Eq. \eqref{G}. This clearly demonstrates that $K_I$ depends on $\lambda$; not being aware of this pre-stretch dependence can induce mistakes.

Finally, Eqs. \eqref{phi_decom}, \eqref{SOLexpansion} and \eqref{biaxial} can be used to calculate the shape of the tip (often called crack tip opening displacement/profile) as $\varphi_x(r,\pi)\!=\!-\kappa(v, \lambda)\varphi_y^2(r,\pi)$, where the tip curvature reads
\begin{equation}
\kappa(v,\lambda)=\left(\frac{\mu \sqrt{2\pi}  D(v, \lambda)}{2\,\alpha_d(v,\lambda)(\alpha_s^2 -1)} \right)^2 \frac{\lambda}{K_I^2} \ .
\label{tip_curvature}
\end{equation}
Since the tip curvature $\kappa$ is, in principle, a directly measurable quantity, the last result can be used to extract the stress-intensity-factor $K_I$. Again, we see that not being aware of the $\lambda$-dependence (e.g. using instead the $\lambda\!=\!1$ result) will lead to mistakes. More generally, analysis of the solution in Eqs. \eqref{biaxial}-\eqref{tip_curvature} reveals that the pre-stretch $\lambda$ has a marked effect on various important physical quantities, and that this effect increases significantly with increasing propagation velocity $v$.

Up to now we considered the asymptotic solution to linear order in $\B {\C U}$, which provided us with some insight into what kind of effects can be associated with the pre-stretch (related points were made in \cite{Guz.1999, Brock_Hanson}). We know, however, that even for small background deformation second order nonlinearities are essential \cite{Bouchbinder.08a,Livne.08,Bouchbinder.09,Autonomy,Bouchbinder.2014}. Consequently, in the next subsection we consider the solution for both $\B {\C U}^{(1)}$ and $\B {\C U}^{(2)}$ under uniaxial background stretch.

\subsection{Semi-analytic second order asymptotic solution}

Here we focus on pre-stressing/straining of the form $\lambda_x\!=\lambda^{-1/2}$ and $\!\lambda_y\!=\!\lambda$, i.e. on uniaxial stretch, which is a more commonly used experimental loading configuration (note that the out-of-plane pre-stretch is $\lambda_z\!=\!\lambda^{-1/2}$, ensuring incompressibility, $\lambda_x \lambda_y \lambda_z\!=\!1$). It corresponds to a pre-stress $s_{yy}^{(0)}\!=\!(\lambda\!-\!\lambda^{-2})\mu$, where the other components of ${\B s}^{(0)}$ vanish. Following the same procedure as above, Eq.~\eqref{1st_2nd_a} takes the form
\begin{eqnarray}
\label{1st_uniaxial}
\hspace{-0.7cm}&&4\pa_{xx} {\C U}_x^{(1)} \!+\! 3\lambda^{-3/2} \pa_{xy} {\C U}_y^{(1)} \!+\! \pa_{yy} {\C U}_x^{(1)}\!=\!\frac{v^2}{c_s^2} \pa_{xx} {\C U}_x^{(1)}\ ,\!\!\!\\
\hspace{-0.7cm}&&\pa_{xx} {\C U}_y^{(1)} \!+\! 3\lambda^{-3/2} \pa_{xy} {\C U}_x^{(1)} \!+\! (1\!+\!3\lambda^{-3}) \pa_{yy} {\C U}_y^{(1)}\!=\!\frac{v^2}{c_s^2}\pa_{xx} {\C U}_y^{(1)} \nonumber,
\end{eqnarray}
with the following boundary conditions at $\theta\!=\!\pm\pi$
\begin{eqnarray}
&\mu \,\pa_y {\C U}_x^{(1)} + \lambda^{-3/2}\mu\,\pa_x {\C U}_y^{(1)} = 0\ ,\nonumber\\
&(1+ 3\lambda^{-3})\mu\,\pa_{y} {\C U}_y^{(1)} + 2\lambda^{-3/2}\mu\,\pa_x {\C U}_x^{(1)} = 0 \ .
\label{1st_uniaxialBC}
\end{eqnarray}

Equations \eqref{1st_uniaxial}-\eqref{1st_uniaxialBC}, which manifestly exhibit elastic anisotropy, can in principle be solved analytically in the asymptotic regime of small $r$. The solution, however, is rather lengthy and we present here instead a semi-analytic procedure to obtain it. To leading order in small $r$, we expect $\B {\C U}^{(1)}\!\sim\!\sqrt{r}$ and the angular dependence to be expressed as a half-integer Fourier series. The sub-leading term in small $r$ (i.e. in the expansion in space), which has not been discussed up to now and which will be included below to enable direct comparison with the experiments to follow, makes a contribution $\propto\!r \cos\theta$ to ${\C U}_x^{(1)}$ and $\propto\!r \sin\theta$ to ${\C U}_y^{(1)}$ (the boundary conditions determine the ratio between the amplitudes).

Therefore, we have
\begin{eqnarray}
{\C U}_x^{(1)}(r,\theta) &=& {\C U}_0 + \bar{K}_I \sqrt{\frac{r\,\Gamma}{\mu}} \sum_{n=1}^N a_n \cos\left[\frac{(2n-1) \theta}{2} \right]\nonumber\\
&+&\frac{T}{12\mu} (1+3\lambda^{-3})\, r \cos\theta \ ,\nonumber\\
{\C U}_y^{(1)}(r,\theta) &=& \bar{K}_I \sqrt{\frac{r\,\Gamma}{\mu}} \sum_{n=1}^N b_n \sin\left[\frac{(2n-1) \theta}{2} \right]\nonumber\\
&-&\frac{T}{6\mu}\lambda^{-3/2}\, r \sin\theta\ .
\label{1st_semi-analytic}
\end{eqnarray}
Here ${\C U}_0$ is a constant (${\C U}_y^{(1)}$ does not include such a constant due to the mode-I symmetry), $\bar{K}_I$ is a dimensionless stress-intensity-factor and $\{a_n, b_n\}$ are dimensionless coefficients. The coefficients of the sub-leading term have been chosen so as to satisfy the boundary conditions of Eq. \eqref{1st_uniaxialBC}. Note that in the limit $\lambda\!\to\!1$ this term corresponds to the so-called $T$-stress and consequently we have chosen the yet undetermined amplitude to agree with the standard result in this limit, where $T$ is a quantity of stress dimensions \cite{Lawn.93,Freund.90,99Bro}.

The coefficients $\{a_n, b_n\}_{n=1-N}$ are determined by the set of linear algebraic equations obtained upon substitution of Eqs. \eqref{1st_semi-analytic} in Eqs. \eqref{1st_uniaxial}-\eqref{1st_uniaxialBC}, where $N$ is chosen to be sufficiently large to ensure convergence. $\bar{K}_I$ is determined by evaluating the J-integral in Eq. \eqref{Gint} (with the linear elastic approximation of $U(\B F)$) and equating $G(v)$ to the fracture energy $\Gamma(v)$. $T$ will be extracted from experimental data.

As stated above, previous work has conclusively demonstrated that second order nonlinearities are important \cite{Bouchbinder.08a,Livne.08,Bouchbinder.09,Autonomy,Bouchbinder.2014}. Hence we wish to calculate $\B {\C U}^{(2)}$. To that aim, we follow the procedure described above to calculate $\B s^{(2)}$, from which Eq.~\eqref{1st_2nd_b} can be obtained. A very detailed, step-by-step, explanation of the mathematical procedure can be found in section 4.2 of \cite{Bouchbinder.2014} and in \cite{Bouchbinder.2009a}. Based on the solution obtained in the framework of the weakly nonlinear theory of fracture \cite{Bouchbinder.08a,Livne.08,Bouchbinder.09,Autonomy,Bouchbinder.2014}, we expect the solution for $\B {\C U}^{(2)}$ to take the form
\begin{eqnarray}
{\C U}_x^{(2)}(r,\theta) &=&  \frac{\Gamma}{\mu}\left(c_0\log{r}+ \sum_{n=1}^N c_n \cos\left[n \theta \right]\right) \ ,\nonumber\\
{\C U}_y^{(2)}(r,\theta) &=&  \frac{\Gamma}{\mu}\left(d_0\,\theta + \sum_{n=1}^N d_n \sin\left[n \theta \right]\right)\ .
\label{2nd_semi-analytic}
\end{eqnarray}
Note that, in principle, the argument of $\log{r}$ should have been made non-dimensional, but this would simply redefine ${\C U}_0$ in Eq. \eqref{1st_semi-analytic} and hence is not essential.

The solution in Eq. \eqref{2nd_semi-analytic} has the property that $\nabla\!_{\B x}{\B {\C U}}^{(2)}\!\sim\!1/r$. As was shown previously, this singularity is special in the sense that it can produce a spurious force in the crack parallel direction \cite{Bouchbinder.09, Autonomy,Rice.74}. To eliminate it, we supplement the equations of motion and boundary conditions with the additional constraint
\begin{equation}
f_x=\int_{-\pi}^{\pi}\!\left[s^{(2)}_{xx}(r,\theta)\cos{\theta}+s_{xy}^{(2)}(r,\theta)\sin{\theta} \right]r\,d\theta\!=\!0 \ ,
\end{equation}
where $f_x$ is the net force per unit sample thickness acting in the $x$ direction on a line of radius $r$ encircling the  crack tip \cite{Bouchbinder.09, Autonomy}.

By satisfying all of these equations, we can calculate the coefficients $\{c_n, d_n\}_{n=0-N}$, where $N$ is chosen to be sufficiently large to ensure convergence. Note that the equation for $\B {\C U}^{(2)}$, cf. Eq.~\eqref{1st_2nd_b}, as well as the boundary conditions, require the knowledge of $\B {\C U}^{(1)}$. Once done, the solution in Eq. \eqref{SOLexpansion}, in the asymptotic region of small $r$, is at hand. In the limit $\lambda\!\to\!1$, the weakly nonlinear theory of fracture is recovered. The next step will be to quantitatively test the predictions of the theory developed above against direct experimental measurements at various values of $\lambda$ and $v$.

\section{Comparison to experiments}

Our goal here is to experimentally test the new theoretical framework. In particular, we would like to compare the predictions of the weakly nonlinear theory (obtained in the limit $\lambda\!\to\!1$ in the theory above) to the predictions of the moderately large deformation theory (which is formulated relative to a finite $\lambda$), against experimental data. To that aim, we conducted experiments with polyacrylamide gels, which are transparent, homogeneous, brittle and incompressible elastomers. The gel composition used here is $14\%$ acrylamide/bis-acrylamide with a $2.7\%$ cross-linker concentration, providing a shear modulus $\mu\!=\!32.3 \pm 1.6$KPa and a shear wave-speed $c_s\!=\!5.6 \pm 0.15$m/sec. $\mu$ is measured prior to each experiment to mitigate any small variations of the gel properties. The energy functional in Eq. \eqref{NH} quantitatively describes the gel \cite{Livne.2010}.

The typical dimensions of our samples are ($x \times y \times z$) $120 \times 120 \times 0.3$mm in the crack propagation, tensile loading and thickness directions, respectively. The thickness was chosen to statistically suppress micro-branching \cite{Livne.07}; while micro-branching can occur for {\em all} velocities $0.3c_s\!<\!v\!<\!0.9c_s$, the probability of exciting them decreases for both thin samples and increased crack accelerations. All of the results presented are for single-crack states in which micro-branching is not observed.

\begin{figure}[here]
\includegraphics[width=0.42\textwidth]{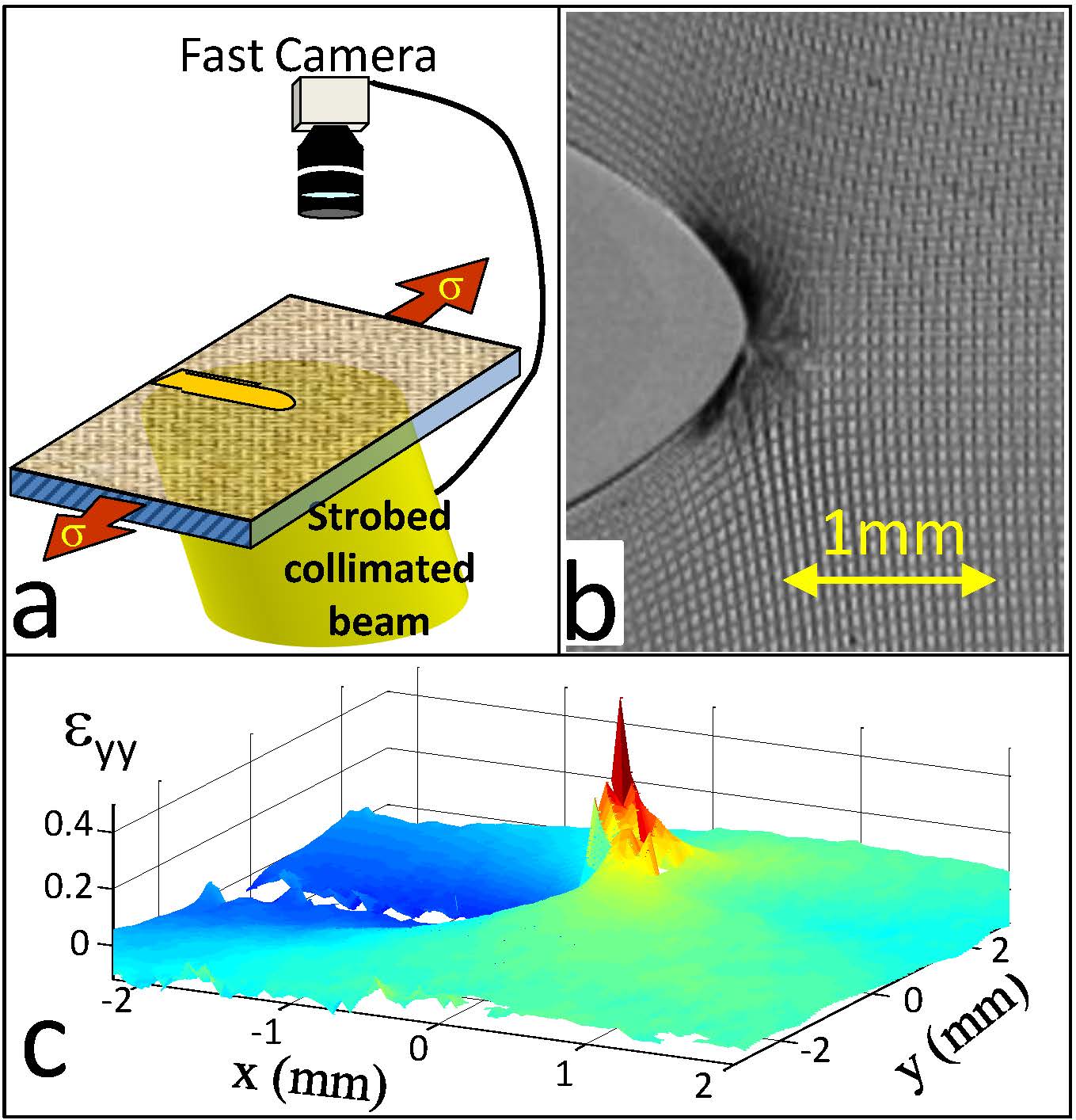}
\caption{(a) Experiments are performed with transparent thin sheets of a brittle elastomer gel with an imprinted grid on one $xy$ face ($y$ is the tensile loading direction and $x$ is the crack propagation direction). Collimated light passed through the sample enables shadowgraph visualization of the grid while pictures are taken with a fast camera. (b) Typical photograph of a crack propagating at $v\!=\!0.53c_s$ through the grid. (c) The measured displacement-gradient field component~$\varepsilon_{yy}(x,y)\!=\!\pa_y u_y(x,y)$ extracted from panel (b).
}\label{fig2}
\end{figure}

The experiments are performed under uniaxial tensile loading in the $y$-direction in accordance with the uniaxial theory presented in the last section. The sample was held at a constant stretch in the range $\lambda\!=\!1.058\!-\!1.129$ prior to crack initiation at the mid-edge of the sample's vertical boundary, as described in Fig.~\ref{fig2}a. Measurements of the crack and its surrounding displacement fields were made with a fast camera (IDT-Y7) focused on an area of $17.4 \times 9.8$mm with a spatial resolution of $1920 \times 1080$ pixels and frame rates of $7400-8100$ frames/sec. The crack velocity for each measurement was set essentially by varying both the imposed strain and position of the measurement area used. All of the cracks are accelerating in accordance with the initial strain imposed by displacing the vertical boundaries of the sample. We achieved desired velocity ranges in each experiment by varying the location of the measurement area according to the initially imposed strain. Due to the small size of the measurement area, the crack velocities were approximately constant throughout the measurement region.

As in \cite{Goldman.15, Prevost.13}, the gels are cast in a mold upon which a rectangular grid is printed on one of its $xy$ surfaces. The grid was formed by lithographic printing on a spin-coated epoxy layer. This process created a perfect square mesh of depth $2\mu$m (in the $z$-direction) and lattice spacing $60\mu$m (in the $xy$ plane). Upon casting, this grid is imprinted on one face of the gel sheets as shown in Fig.~\ref{fig2}b.
Shadowgraphy, using strobed lighting ($2\mu$sec duration), is used to image both the deformed grid and crack opening profile as a crack's tip propagates across the field of view (Fig. \ref{fig2}a). The location of the center of each grid point in the deformed grid is determined to within 10$\mu$m resolution. The displacement fields were acquired by comparing the position of the grid points under deformation to their position in a deformation-free system. Fig.~\ref{fig2}c demonstrates a typical measurement of the displacement-gradient field (here the $\varepsilon_{yy}\!=\!\pa_y u_y$ component is shown).

\begin{figure}[here]
\includegraphics[width=0.465\textwidth]{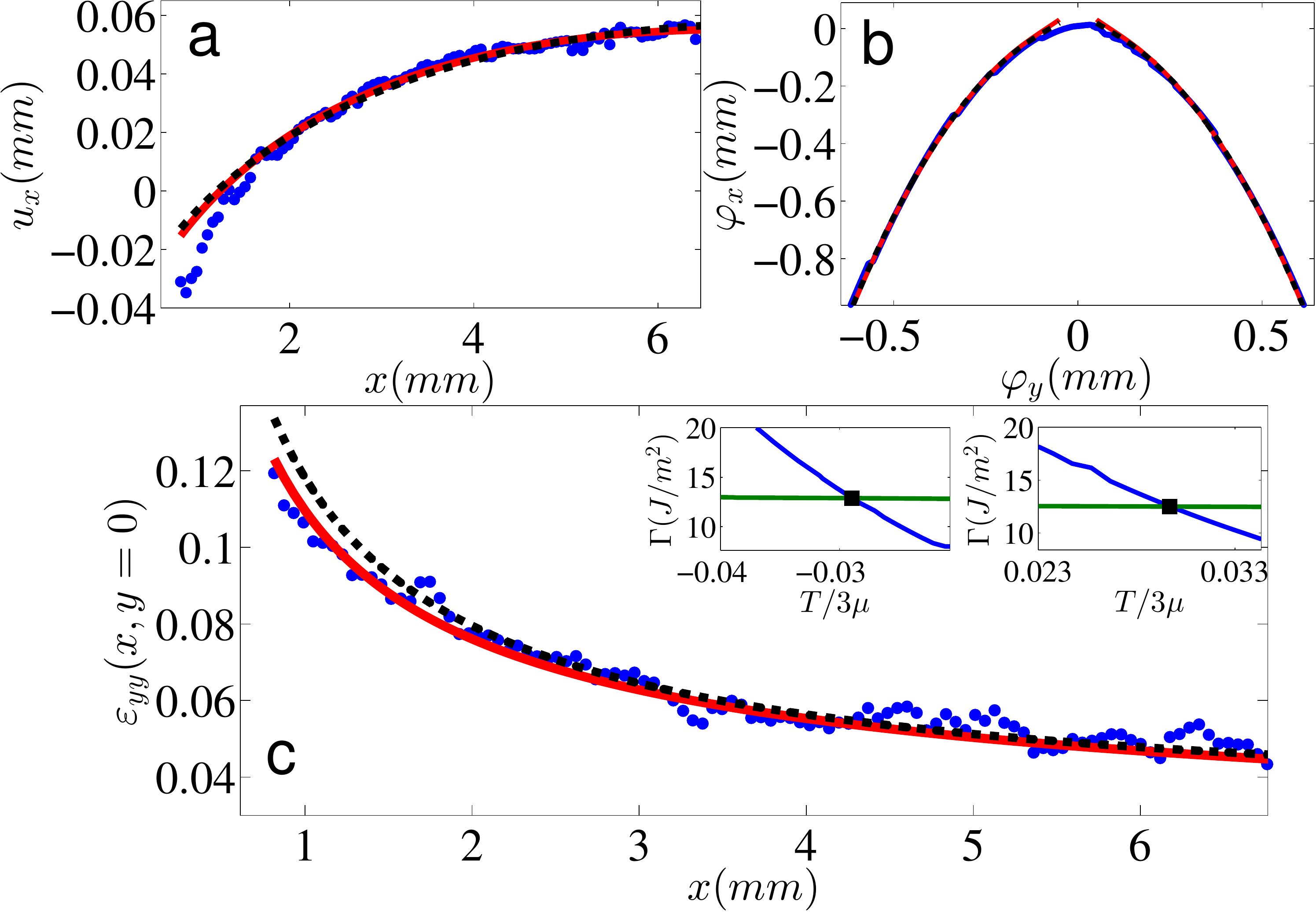}
\caption{Measurements of a crack propagating at $v\!=\!0.29 c_s$ under uniaxial stretch of $\lambda\!=\!1.058$. (a) The measured displacement in the crack parallel direction ahead of the tip $u_x(r\!=\!x,\theta\!=\!0)$ (blue circles). Fits to the weakly nonlinear theory (red solid line) and to the moderately large deformation theory (black dashed line) are superimposed. (b) The measured crack tip opening profile $\varphi_x(r,\theta\!=\!\pi)\left[\varphi_y(r,\theta\!=\!\pm\pi)\right]$ (blue solid line). As in panel (a), the fits to the two theories are superimposed. The fitting parameters $\{{\C U}_0, T, \Gamma\}$ were obtained for each theory, where the convergence of the iterative procedure in the $\Gamma-T$ parametric plane is shown in the inset of panel (c), for the weakly nonlinear theory (left) and moderately large deformation theory (right). The intersections in the $\Gamma-T$ plane, denoted by black squares, are the values chosen by our fitting procedure (see text). In addition, we obtained ${\C U}_0\!\simeq\!25\mu$m for both theories. Note that the green and blue curves were obtained by the fits to the crack opening profile and $u_x(x,y\!=\!0)$, respectively. (c) The measured tensile strain ahead of the tip $\varepsilon_{yy}(x,y\!=\!0)\!=\!\pa_y u_y(x,y)|_{y=0}$ (blue circles). The predictions of the two theories,  using the parameters obtained by the fits displayed in panels (a) and (b), are superimposed (lines as above).}
\label{fig3}
\end{figure}

In Fig. \ref{fig3} we compare the experimental measurements with both the weakly nonlinear theory and the moderately large deformation theory for a moderate stretch $\lambda\!=\!1.058$ and crack propagation velocity $v\!=\!0.29c_s$. The basic field of interest is the displacement field $\B u(x,y)$, which is measured directly and is theoretically obtained from Eq. \eqref{SOLexpansion} through $u_x(x,y)\!=\!(\lambda^{-1/2}-1)\,x + {\C U}_x(x,y)$ and $u_y(x,y)\!=\!(\lambda-1)\,y + {\C U}_y(x,y)$. Each theory contains three parameters~$\{{\C U}_0, T, \Gamma\}$ that are not determined by the asymptotic analysis. ${\C U}_0$ and $ T$ correspond, respectively, to a small constant shift of the crack tip location and to the $T$-stress. As our measurements are not solely within the asymptotic (singular) region, both of these quantities are needed for a good quantitative comparison.

For both theories, we determined the values of $\{{\C U}_0, T, \Gamma\}$ by the following procedure. We considered both $u_x(r\!=\!x,\theta\!=\!0)$ and $\varphi_x(r,\theta\!=\!\pi)\left[\varphi_y(r,\theta\!=\!\pm\pi)\right]$ (in the latter, $r$ parameterizes the function), i.e. the crack parallel displacement ahead of the tip and the crack tip opening profile, respectively. The first function is quite sensitive to $T$, whereas the latter is very sensitive to $\Gamma$. We iteratively performed a 3-parameter fit over these two functions until the best fit with the {\em same} $T$ and $\Gamma$ is obtained. ${\C U}_0$ is not constrained to be the same, but turns out to be so. The resulting fits for $u_x(r\!=\!x,\theta\!=\!0)$ and $\varphi_x(r,\theta\!=\!\pi)\left[\varphi_y(r,\theta\!=\!\pm\pi)\right]$ are shown in Figs. \ref{fig3}a-b, respectively. The convergence in the $\Gamma-T$ parametric plane, for both the weakly nonlinear and  moderately large deformation theories, is demonstrated in the inset of Fig. \ref{fig3}c.

Figures \ref{fig3}a-b suggest that at this level of pre-stressing/straining (relatively low in the context of the results to follow) and crack propagation velocity the two theories appear to be almost indistinguishable, at least as far as the quantities shown are considered. The parameters, however, are not the same. In particular, as the inset of Fig. \ref{fig3}c clearly demonstrates, in this case $\Gamma$ is quite similar while $T$ is not. The latter difference is expected since the background stretches in the moderately large deformation theory, cf. Eq. \eqref{phi_decom}, directly affect the $T$-stress term and actually make $T$ positive (while it is negative in the weakly nonlinear theory). We emphasize, though, that when considered relative to the undeformed configuration, the $T$-stress in the moderately large deformation theory is also negative (as is common for uniaxial loading).

Once $\{{\C U}_0, T, \Gamma\}$ are determined, there are no longer any free parameters and the two theoretical frameworks can be independently tested against other experimentally measured quantities. In particular, we will use two types of tests:\\
$\bullet\,\,$Comparing the predictions of the two theories to the measured tensile strain ahead of the tip, $\varepsilon_{yy}(x,y\!=\!0)\!=\!\pa_y u_y(x,y)|_{y=0}$.\\
$\bullet\,\,$Comparing the predictions of the two theories for the fracture energy to $\Gamma(v)\!=\!G(v)\!=\!J/v$ independently obtained from the J-integral in Eq. \eqref{Gint} (using the measured $\B\varphi(x,y)$, as in \cite{Livne.2010}).
\begin{figure}[here]
\includegraphics[width=0.48\textwidth]{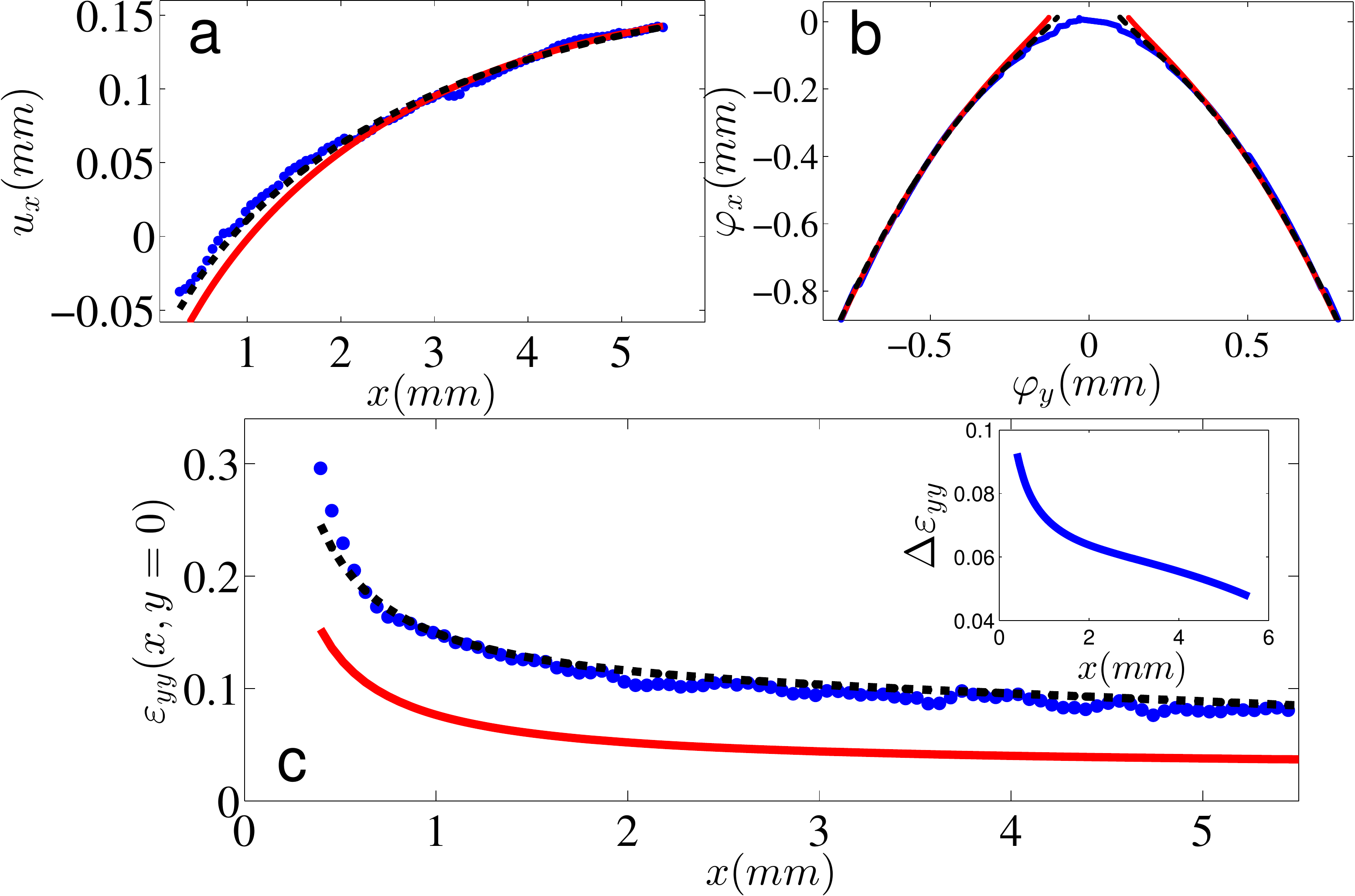}
\caption{Measurements of a crack propagating at $v\!=\!0.74c_s$ under uniaxial strain of $\lambda\!=\!1.096$, where everything is as in Fig. \ref{fig3}, except for the inset of panel (c). In the inset, we plot the difference between predictions of the moderately large deformation and weakly nonlinear theories for $\varepsilon_{yy}(x,y\!=\!0)$ (shown in the main panel), denoted by $\Delta\varepsilon_{yy}$. The difference is a nontrivial spatially-varying function that increases significantly as the crack tip is approached.}
\label{fig4}
\end{figure}

The results of these parameter-free comparisons are shown in Fig. \ref{fig3}c (main panel) and Fig. \ref{fig6} (focus on $v\!=\!0.29c_s$). Figure \ref{fig3}c shows that the predictions of the two theories are in good agreement with the measured data, where the weakly nonlinear theory is doing slightly better. Figure \ref{fig6} shows that for $v\!=\!0.29c_s$, the fracture energies $\Gamma$ predicted by the two theories are in agreement with the one independently calculated through the J-integral. All in all, we conclude that for relatively low levels of pre-stressing/straining and crack propagation velocities, the two theories appear rather consistent with one another and quantitatively agree with the experiments. Furthermore, this analysis reconfirms the validity of the weakly nonlinear theory, as reported previously \cite{Bouchbinder.08a,Livne.08,Bouchbinder.09,Autonomy,Bouchbinder.2014}. The main question now is what happens as the pre-stress/strain and the crack propagation velocity are significantly increased.

In Fig. \ref{fig4} we repeat the analysis presented in Fig. \ref{fig3} for a crack propagating at $v\!=\!0.74c_s$ under significantly increased pre-stressing/straining corresponding to $\lambda\!=\!1.096$. Figures \ref{fig4}a-b indicate that both theories can be reasonably fitted to the measured $u_x(r\!=\!x,\theta\!=\!0)$ and $\varphi_x(r,\theta\!=\!\pi)\left[\varphi_y(r,\theta\!=\!\pm\pi)\right]$, where the moderately large theory is doing better with respect to the former. The parameter-free comparison shown in Fig. \ref{fig4}c, however, reveals a striking difference between the two theories; the prediction of the moderately large deformation theory for $\varepsilon_{yy}(x,y\!=\!0)$ is significantly better than the prediction of the weakly nonlinear theory and is in good quantitatively agreement with the measurements. Furthermore, Fig. \ref{fig6} shows that the moderately large deformation theory predicts a fracture energy $\Gamma$ much closer to the independent J-integral estimate than the weakly nonlinear theory (focus on $v\!=\!0.74c_s$).
\begin{figure}[here]
\includegraphics[width=0.465\textwidth]{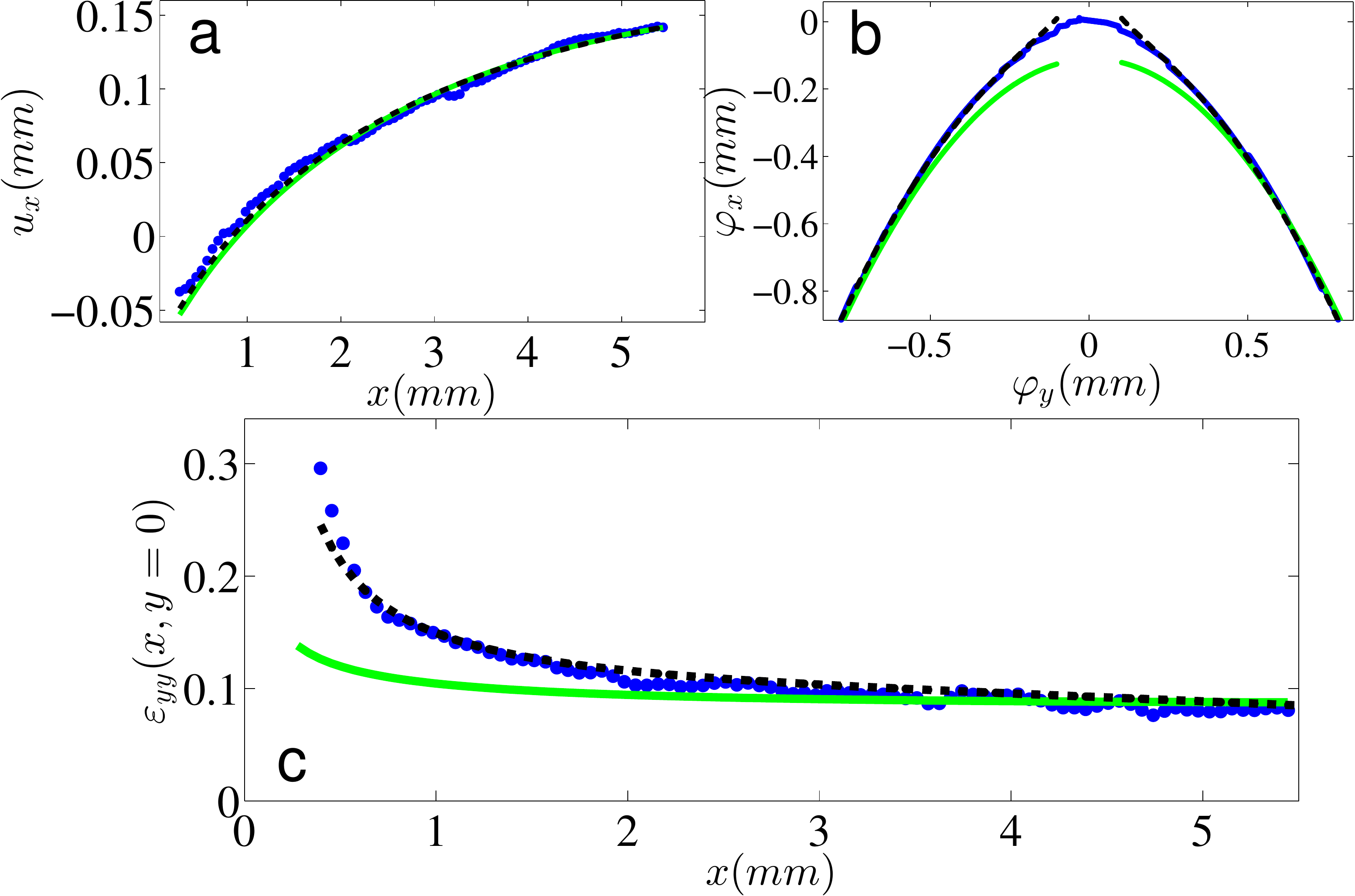}
\caption{The same as Fig.~\ref{fig4}, except that the weakly nonlinear analysis is replaced with a moderately large deformation analysis truncated to first order (green solid line). For the latter, we follow the same fitting procedure as before, but set $\B {\C U}^{(2)}\!=\!0$. The best fitting parameters are ${\C U}_0\!=\!100\mu$m, $T/3\mu\!=\!0.01$ and $\Gamma\!=\!19.8$ J/m$^2$. The necessity of the second order nonlinearities is evident from the panels (b) and (c), where significant discrepancies between the first order truncated moderately large deformation theory and the experimental data are observed.}
\label{fig5}
\end{figure}

It is crucial to understand that the pre-stretching $\lambda$ affects all physically relevant quantities in the problem, as in the analytic example of Sec. \ref{subsec:analytic}, in a nontrivial way. For example, in the inset of Fig. \ref{fig4}c we show the difference between the predictions of the two theories shown in the main panel, which is a nontrivial spatially-varying function that increases significantly as the crack tip is approached. The quantitative analysis presented in Figs. \ref{fig4} and \ref{fig6} has been repeated for many cracks with propagation velocities in the range $v\!=\!0.26c_s-0.75 c_s$ and pre-stretching levels in the range $\lambda\!=\!1.058-1.129$ (since the cracks are mildly accelerating, each pre-stretch produces a range of crack velocities). As in Fig. \ref{fig4}c, at stretches of about $\lambda\!=\!1.1$ or higher and large propagation velocities, the moderately large deformation theory predicts $\varepsilon_{yy}(x,y\!=\!0)$ significantly better than the weakly nonlinear theory.

Before we discuss the predictions for fracture energy $\Gamma(v)$, we briefly highlight the importance of second order nonlinearities in the expansion relative to the pre-stretched configuration. This has been previously established in relation to the weakly nonlinear theory, cf. Fig. 1 in~\cite{Bouchbinder.08a}. To show that this remains valid in the case of the moderately large deformation theory, we plot in Fig.~\ref{fig5} everything as in~Fig.\ref{fig4}, except that the weakly nonlinear analysis is replaced with a moderately large deformation analysis truncated to first order. For the latter, we follow the same fitting procedure for $\{{\C U}_0, T, \Gamma\}$ as before, but set $\B {\C U}^{(2)}\!=\!0$. The results clearly demonstrate that second order nonlinearities are indeed essential. In particular, the footprints of the missing logarithmic term (which appears in ${\C U}_x^{(2)}$, cf. Eq.~\eqref{2nd_semi-analytic}) and the stronger singularity ($\nabla\!_{\B x}\B {\C U}^{(2)}\!\sim\!1/r$) are evident in Figs.~\ref{fig5}b-c. The rather dramatic failure of the linear order moderately large deformation theory to predict $\pa_u u_y(x,y\!=\!0)$ is related to the properties of the LEFM asymptotic fields, which predict a negative $\pa_u u_y(x,y\!=\!0)$ for sufficiently high velocities \cite{Bouchbinder.08a}.

Figure \ref{fig6} shows $\Gamma(v)$ for the full range of pre-stretches and propagation velocities considered in this study, indicating that the moderately large deformation theory predicts fracture energy values closer to the independent J-integral estimate than the weakly nonlinear theory (accurately predicting the J-integral values up to $\sim\!10\%$). We would like to stress that the performed fits are robust. In fact, we checked that using $\varepsilon_{yy}(x,y\!=\!0)$ and $\varphi_x(r,\theta\!=\!\pi)\left[\varphi_y(r,\theta\!=\!\pm\pi)\right]$ to determine $\{{\C U}_0, T, \Gamma\}$ and then testing the predictions for $u_x(r\!=\!x,\theta\!=\!0)$ and $\Gamma$ (independently obtained through the J-integral) yielded similar results to those reported above.
\begin{figure}[here]
\includegraphics[width=0.47\textwidth]{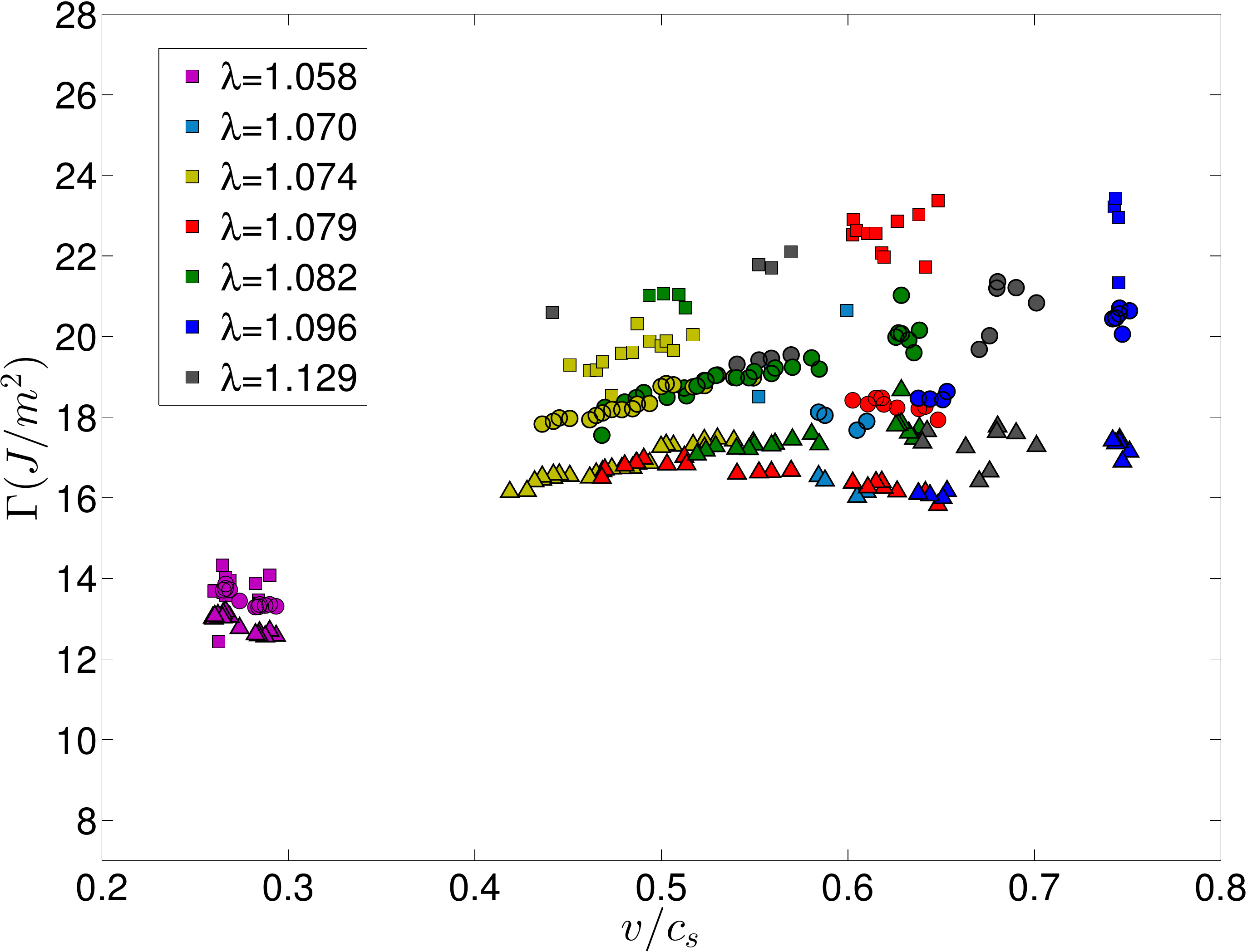}
\caption{The measured fracture energy $\Gamma(v)$ (squares) with propagation velocities in the range $v\!=\!0.26c_s-0.75 c_s$ and pre-stretching levels in the range $\lambda\!=\!1.058-1.129$ (different colors correspond to different $\lambda$, cf. legend). $\Gamma(v)$ was estimated from the J-integral of Eq. \eqref{Gint}, using the measured $\B\varphi(x,y)$. The predictions of the weakly nonlinear theory (triangles) and of the  moderately large theory (circles) are presented for comparison.}
\label{fig6}
\end{figure}

Taken together, we believe that the quantitative analysis presented here clearly shows that the moderately large deformation framework offers significantly better approximations to the direct measurements than the weakly nonlinear theory at moderately large levels of external loadings and high propagation velocities.

\section{Concluding remarks}
\label{summary}

In this paper we developed and experimentally tested a dynamic fracture theory of highly-deformable materials which fail under the application of large external strains. The theory is based on a {\em second order} expansion in the displacement-gradients with respect to a nonlinearly stretched reference state. While the theory is mathematically well-defined, its physical range of applicability -- and the mere existence of such a range -- are not {\em a-priori} guaranteed. The reason for this is that as the background strains are already rather large, amplification of deformation near the crack tip may render a perturbative approach inappropriate.

Our direct experiments showed firstly that the background strain has a significant effect on the deformation fields surrounding crack tips. Even at moderate strains, the standard theory fails to provide a good description of the near tip fields. Secondly, and quite surprisingly, the experiments showed that the new theory provides a good description of these fields. This central finding emerged because the amplification of deformation in the region of interest was not enormous and due to the inclusion of second order terms in the theory. These results imply that the theory may have a robust range of applicability, at least at moderately large background strains and high propagation velocities, but possibly also at larger background deformation.

This convincing experimental support indicates that the proposed theoretical development offers a framework to understand the dynamic fracture of soft materials that fail under large pre-stressing/straining (of the order of $\sim\!10-20\%$ or larger), going significantly beyond the standard fracture theory of ordinary materials that fail under strains of $\sim\!1\%$. The theory shows that in order to quantitatively understand the fields that drive material failure near crack tips, the deformation-induced anisotropy and fundamental material properties such as the fracture energy, the pre-stressing/straining needs to be properly taken into account in such materials. The theory may find applications in a range of problems dealing with the failure of soft materials, from food processing to tissue rupture.

It is important to note that while some existing literature exclusively focusses on the inner most asymptotic crack tip region in highly-deformable materials \cite{Knowles.73, 82Steph, LeStumpf.93, 95Geubelle, Long.2011, Tarantino.99, 05Tarantino}, our theory takes into account the pre-strained/stressed large scales and links them to the near tip region as approached from the outside. As such, the theory should be regarded as intermediate asymptotics. The experiments presented above, which quantitatively support the theory, are able to probe this intermediate asymptotic region.

One insight emerging from this work is that the form of these intermediate asymptotic solutions is highly influenced by the magnitude of the background deformations. In the classic LEFM theory, the loading and background stresses are solely accounted for by their influence on the intensity of the universal singularity, the stress intensity factor. Here, we have shown that large strains inherent in the external loading actually influence the fields in a variety of subtle ways that can not solely be accommodated by a change in the value of the stress intensity factor.

Finally, note that various highly-deformable materials fail under background deformation of the order of 100\% or even larger, a regime that has not been probed by the experiments presented above. Therefore, it remains to be seen in future work how far one can push the theoretical framework developed here.\\

{\bf Acknowledgements} E.B. benefitted from discussions with E. Brener. J.F.
and T.G. acknowledge support from the European Research Council (Grant No 267256).
E.B. and J.F. acknowledge support from the James S.
McDonnell Fund (Grant No 220020221). E.B. acknowledges support from the Minerva Foundation with funding
from the Federal German Ministry for Education and
Research, the Harold Perlman Family Foundation and the
William Z. and Eda Bess Novick Young Scientist Fund.

\providecommand*{\mcitethebibliography}{\thebibliography}
\csname @ifundefined\endcsname{endmcitethebibliography}
{\let\endmcitethebibliography\endthebibliography}{}





\end{document}